\documentclass[journal,onecolumn]{IEEEtran}

\usepackage{graphicx,cite}
\usepackage{amsmath}
\usepackage{subfig} 
\usepackage{adjustbox}
\usepackage{multirow}
\usepackage{romannum}
\ifCLASSINFOpdf

\else

\fi

\begin{document}
\title{Through-the-Wall Radar under Electromagnetic Complex Wall: A Deep Learning Approach}

\author{Fardin~Ghorbani and {} Hossein~Soleimani

\thanks{Corresponding Author: Hossein Soleimani (email: hsoleimani@iust.ac.ir)}
\thanks{F. Ghorbani and H. Soleimani are with the School of Electrical
Engineering, Iran University of Science and Technology, Tehran, Iran}
}

\maketitle

\begin{abstract}
This paper employed deep learning to do two-dimensional, multi-target locating in Through-the-Wall Radar under conditions where the wall is treated as a complex electromagnetic medium. We made five assumptions about the wall and two about the number of targets. There are two target modes available: single target and double targets. The wall scenarios include a homogeneous wall, a wall with an air gap, an inhomogeneous wall, an anisotropic wall, and an inhomogeneous-anisotropic wall. Target locating is accomplished through the use of a deep neural network technique. We constructed a dataset using the Python FDTD module and then modeled it using deep learning. Assuming the wall is a complex electromagnetic medium, we achieved $97.7\%$ accuracy for single-target 2D locating and $94.1\%$ accuracy for two-target locating. Additionally, we noticed a loss of $10\%$ to $20\%$ inaccuracy when noise was added at low SNRs, although this decrease dropped to less than $10\%$ at high SNRs.
\end{abstract}


\IEEEpeerreviewmaketitle

\section{Introduction}

\IEEEPARstart{R}{ecently}, Through-the-Wall Imaging (TWRI) has grown in prominence as a field of study. TWR can be used to locate, identify, classify, and track humans and moving objects beyond the wall\cite{kumar2014experimental,buonanno2013new}. Due to the interference present in chaotic and multidimensional environments, it is difficult to identify and classify targets hidden behind walls and inside structures. Electromagnetic waves are used to accomplish this, but advancements in this multifaceted technology require a thorough understanding of the complexities of EM wave interaction with internal and external objects. Non-destructive TWRI technology is effective at detecting and locating invisible targets. There has been a lot of research conducted in this field. In the presence of wall parameters, accurate target identification is possible \cite{ahmad2005synthetic}\cite{ahmad2008three}. Another technique for determining the location of a human is to use doppler\cite{ding2020human,chauhan2020through}. Certain techniques must be investigated in order to compensate for wall effects and obtain precise positioning and high-quality, focused images. As a result, wall parameters such as permittivity, thickness, and conductivity are critical for accurate target locating and imaging. There are two types of methods for estimating wall parameters: conventional methods and machine-based methods. The time-delay approach\cite{protiva2011estimation}, filter-based methods \cite{jin2012image}, M-Sequence sensor, and continuous basis estimator\cite{fereidoony2017m} are all examples of conventional methods. Additionally, machine learning algorithms have been used to estimate wall parameters \cite{zhang2015application,zhang2016efficient}. Another approach for locating targets behind a wall is to employ techniques that do not require knowledge of the wall's parameters. Several algorithms, including inverse linear scattering algorithms based on the first-order Born approximation \cite{soldovieri2007through}, auto-focusing strategies based on the spectrum Greens
function \cite{li2009novel}, auto-focusing techniques based on higher order
statistics \cite{ahmad2007autofocusing}, and the trajectory intersection method for estimating
the target position \cite{wang2006new} have been used. 
Conventional machine learning approaches were recently used for target detection and classification in TWR. 
Bufler et al. classified indoor targets using machine learning techniques \cite{bufler2016radar}. The support vector machine (SVM) is used in this study to distinguish humans from indoor targets. Additionally, researchers have proposed a kernel extreme machine learning technique for locating through walls with unknown parameters \cite{zhang2020real}. Zhang et al. developed a two-dimensional positioning method based on the assumption of a homogeneous wall and a circular metal cylinder target. Additionally, in \cite{zhang2020robust}, an extreme machine learning is used  to propose a 3D locating technique for a homogeneous single layer wall surrounding a spherical metallic target. Wood et al. \cite{wood2020through} utilized a machine learning approach to accomplish three goals, one of which is predicting an object's location. This article performs two-dimensional locating assuming a circular target using the K-Nearest Neighbors (KNN) algorithm and a homogeneous non-magnetic wall. Also \cite{ghorbani2021simultaneous} presents a deep learning-based method to estimate wall parameters and object parameters simultaneously when the wall is homogeneous.

Machine Learning is the study of designing machines that learn from both provided examples and their own experiences. Indeed, machine learning is an attempt to develop a machine capable of learning and functioning without explicitly planning and dictating individual actions through the use of algorithms. Rather than programming each action, machine learning uses data to feed a general algorithm, which then constructs its logic based on the data \cite{bishop2006pattern}. We are witnessing the penetration of machine learning into electromagnetic problems as a result of the increasing development of machine learning and its potential to solve a variety of problems\cite{xu2020deep,sharma2020machine,ghorbani2021deep,ghorbani2021deepp}.
 The majority of machine learning algorithms are capable of detecting and classifying hidden patterns and signals\cite{ghorbani2020eegsig}. Due to this property, machine learning is an excellent tool in the field of radar. We may find acceptable applications for these methods by applying machine learning algorithms to the radars \cite{travassos2020artificial},\cite{majumder2020deep}.
\begin{figure*}[h]
	\centering
	\includegraphics[scale=0.15]{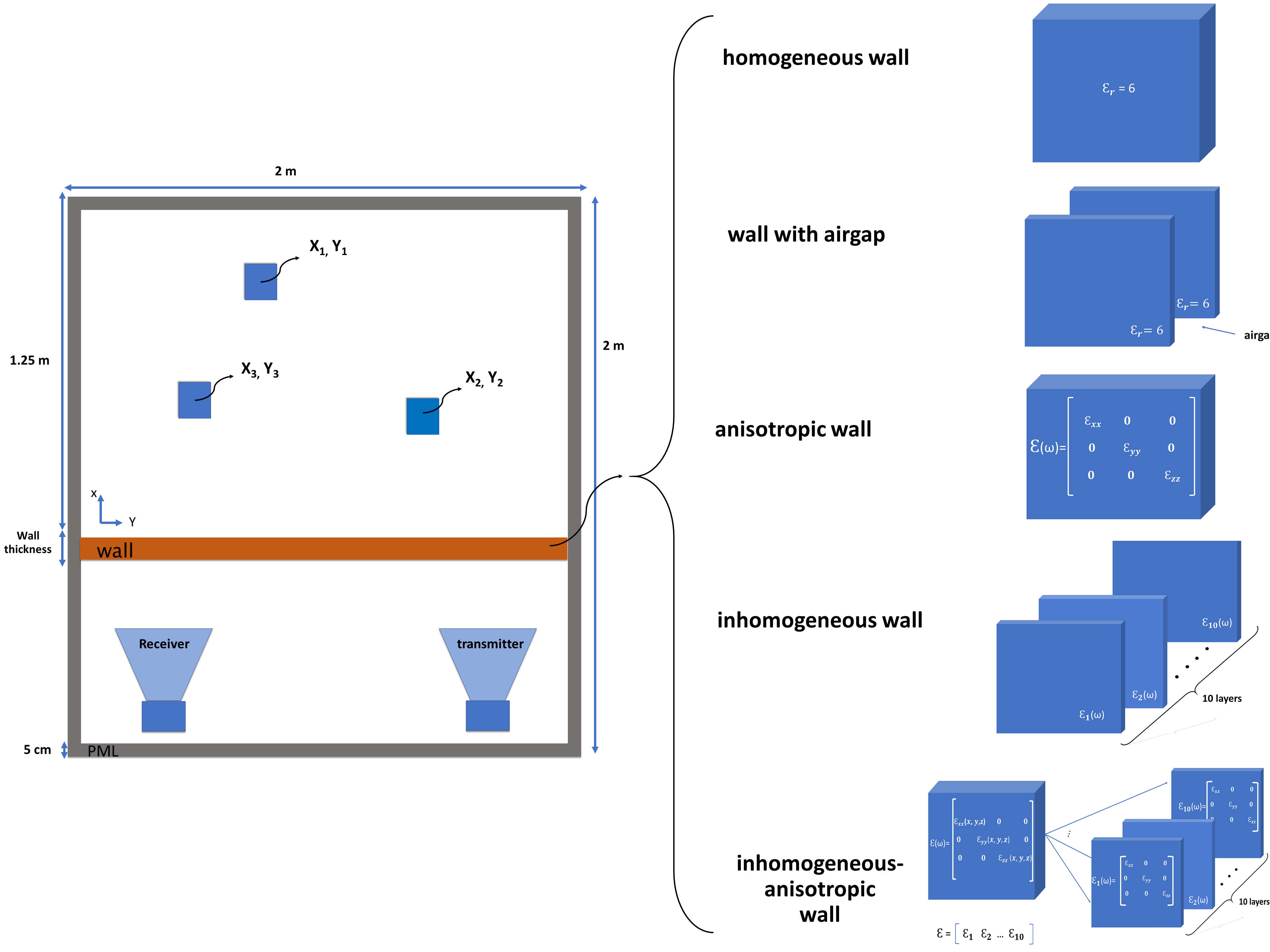}
	\caption{An overview of problem} 
\end{figure*}
\\

The purpose of this paper is to determine the two-dimensional location of targets. We employ multiple targets rather than a single one. Additionally, we use a complex electromagnetic model to create a more accurate representation of the wall, making it more difficult to locate targets. Rather than using standard machine learning techniques, we employ deep learning. We consider five possible scenarios for the wall: homogeneous wall, airgap wall, inhomogeneous wall, anisotropic wall, and inhomogeneous-anisotropic wall. Following that, we conduct multi-target localization.

\section{METHODOLOGIES}

\subsection{Complex Electromagnetic Media}

In this paper, we considered five distinct wall models: homogeneous wall, air gap wall, inhomogeneous wall, anisotropic wall, and inhomogeneous-anisotropic wall. To approximate an inhomogeneous wall, we use multiple homogeneous layers, each with a specified value of $\varepsilon$. When the homogeneous layers are combined, an inhomogeneous wall is created.

\begin{equation}
\begin{cases}\bar{D}(x,y,z, \omega) = \varepsilon (x,y,z,\omega ) \bar{E}(x,y,z, \omega) \\ 
\bar{B}(x,y,z, \omega) = \mu (x,y,z,\omega ) \bar{H}(x,y,z, \omega)\end{cases}
\end{equation}

If the wall is a perfect nonmagnetic dielectric with only one direction of inhomogeneity, the permittivity is given by $\varepsilon (x,\omega)$, $\mu=1$, and $\sigma=0$. The inhomogeneous wall is approximated by a series of homogeneous layers adjacent to one another; for example, a ten-centimeter wall is approximated by ten layers with varying permittivity, the permittivity values of which were $\varepsilon_{r} = [3,6,4,5,6,3,6,3,5,2]$. The wall view is depicted in Figure 2.

\begin{figure}
	\centering
	\includegraphics[scale=0.08]{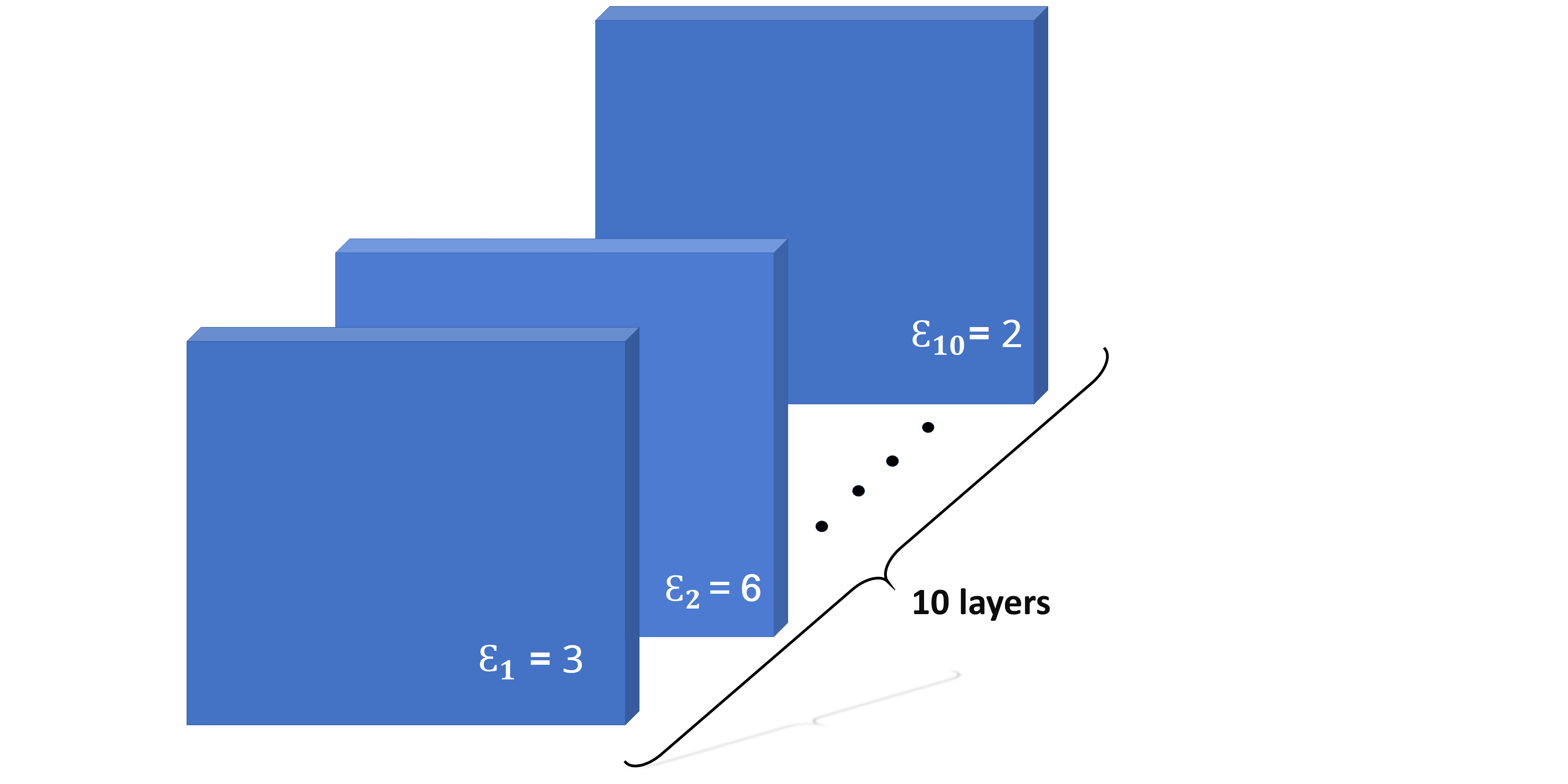}
	\caption{Overview of inhomogeneous wall approximation with homogeneous wall layers approximation} 
\end{figure}

In isotropic environments, the orientation of the environment's electric or magnetic polarization is precisely determined by the direction of the external electric or magnetic field. As a result, the electrical and magnetic permeability coefficients of such environments, $\varepsilon$ and $\mu$, are expressed as numerical coefficients. On the other hand, there are numerous materials whose electrical or magnetic polarization is not oriented in the direction of electric or magnetic fields. As a result, the electrical and magnetic permeability coefficients of these environments, also referred to as anisotropic environments, must be expressed using matrices and tensors. In mathematics, anisotropy is defined as the lack of symmetry with respect to a collection of spatial rotation transformations. An anisotropic structure is one that appears differently from different axes. An anisotropic structure is one that has at least one non-scalar structural parameter in electromagnetism. In other words, in anisotropic environments, field $D(\bar{r, \omega })$ is not in the $E(\bar{r, \omega })$ direction, or field $B(\bar{r, \omega })$ is not in the $H(\bar{r, \omega })$ direction. 
\begin{equation}
\begin{cases}\bar{D}(\bar{r}, \omega ) = \bar{\bar{\varepsilon}} (\omega ) \bar{E}(\bar{r}, \omega ) \\ 
\bar{B}(\bar{r}, \omega ) = \bar{\bar{\mu }}(\omega ) \bar{H}(\bar{r}, \omega )\end{cases}
\end{equation}

Where :

\begin{equation}
\bar{\bar{\varepsilon}} (\omega) = \begin{bmatrix}\varepsilon_{xx} & \varepsilon_{xy} & \varepsilon_{xz} \\
\varepsilon_{yx} & \varepsilon_{yy} & \varepsilon_{yz} \\ 
\varepsilon_{zx} & \varepsilon_{zy} & \varepsilon_{zz} \end{bmatrix}
\end{equation}

\begin{equation}
\bar{\bar{\mu }}(\omega) = \begin{bmatrix}\mu_{xx} & \mu_{xy} & \mu_{xz} \\
\mu_{yx} & \mu_{yy} & \mu_{yz} \\
\mu_{zx} & \mu_{zy} & \mu_{zz} \end{bmatrix}
\end{equation}
We utilize a uni-axial anisotropic non-magnetic perfect dielectric wall in this work, where the non-main diagonal of the $\varepsilon$ matrix is equal to zero, and only the main diagonal has a value of $\mu=1$, $\sigma=0$, and permittivity of the wall is equal to the following tensor.

\begin{equation}
\bar{\bar{\varepsilon}} = \begin{bmatrix}\varepsilon_{xx} & 0 &0 \\
0 & \varepsilon_{yy} &0 \\ 
0 &0 & \varepsilon_{zz} \end{bmatrix} = \begin{bmatrix}6 & 0 &0 \\
0 & 4&0 \\ 
0 &0 &2 \end{bmatrix}
\end{equation}

Another model for the wall is inhomogeneous-anisotropic. We represent the wall as multiple homogeneous layers, with each layer having a distinct tensor $\varepsilon$ for inhomogeneity and anisotropy. Indeed, we have a wall composed of many homogeneous-anisotropic layers stacked on top of one another to form an inhomogeneous-anisotropic wall. The following is the problem's assumed permittivity:

\begin{align}
\begin{split}
\bar{\bar{\varepsilon}} (x,\omega)&= \begin{bmatrix}\varepsilon_{xx}(x) & 0 &0 \\
0 & \varepsilon_{yy}(x) &0 \\ 
0 &0 & \varepsilon_{zz}(x) \end{bmatrix}
\\& = \begin{bmatrix} 
\bar{\bar{\varepsilon_{1}}},\bar{\bar{\varepsilon_{2}}},... ,\bar{\bar{\varepsilon_{10}}} \end{bmatrix}
\end{split}
\end{align}

 For the sake of simplicity, let $\varepsilon=[\varepsilon_{xx},\varepsilon_{yy},\varepsilon_{zz}]$ in Equation 5. In this case, we assumed is ten centimeter thick and approximated it with ten distinct layers. The permittivity values of the layers have been assumed to be those in Table \Romannum{1}. The overview of the inhomogeneous-anisotropic wall is shown in Figure 3.

\begin{table}[h]
	\centering
	\caption{The values of permittivity in the approximation of inhomogeneous-anisotropic wall}
	\begin{tabular}{|l|l|}
		\hline
		layer number & $\varepsilon= [\varepsilon_{xx},\varepsilon_{yy},\varepsilon_{zz}]$         \\ \hline
		1            & {[}6,3,2{]} \\ \hline
		2            & {[}5,5,2{]} \\ \hline
		3            & {[}6,4,2{]} \\ \hline
		4            & {[}4,6,2{]} \\ \hline
		5            & {[}3,4,2{]} \\ \hline
		6            & {[}2,3,2{]} \\ \hline
		7            & {[}5,2,2{]} \\ \hline
		8            & {[}2,4,2{]} \\ \hline
		9            & {[}4,3,2{]} \\ \hline
		10           & {[}3,5,2{]} \\ \hline
	\end{tabular}
\end{table}

\begin{figure}[h]
	\centering
	\includegraphics[scale=0.08]{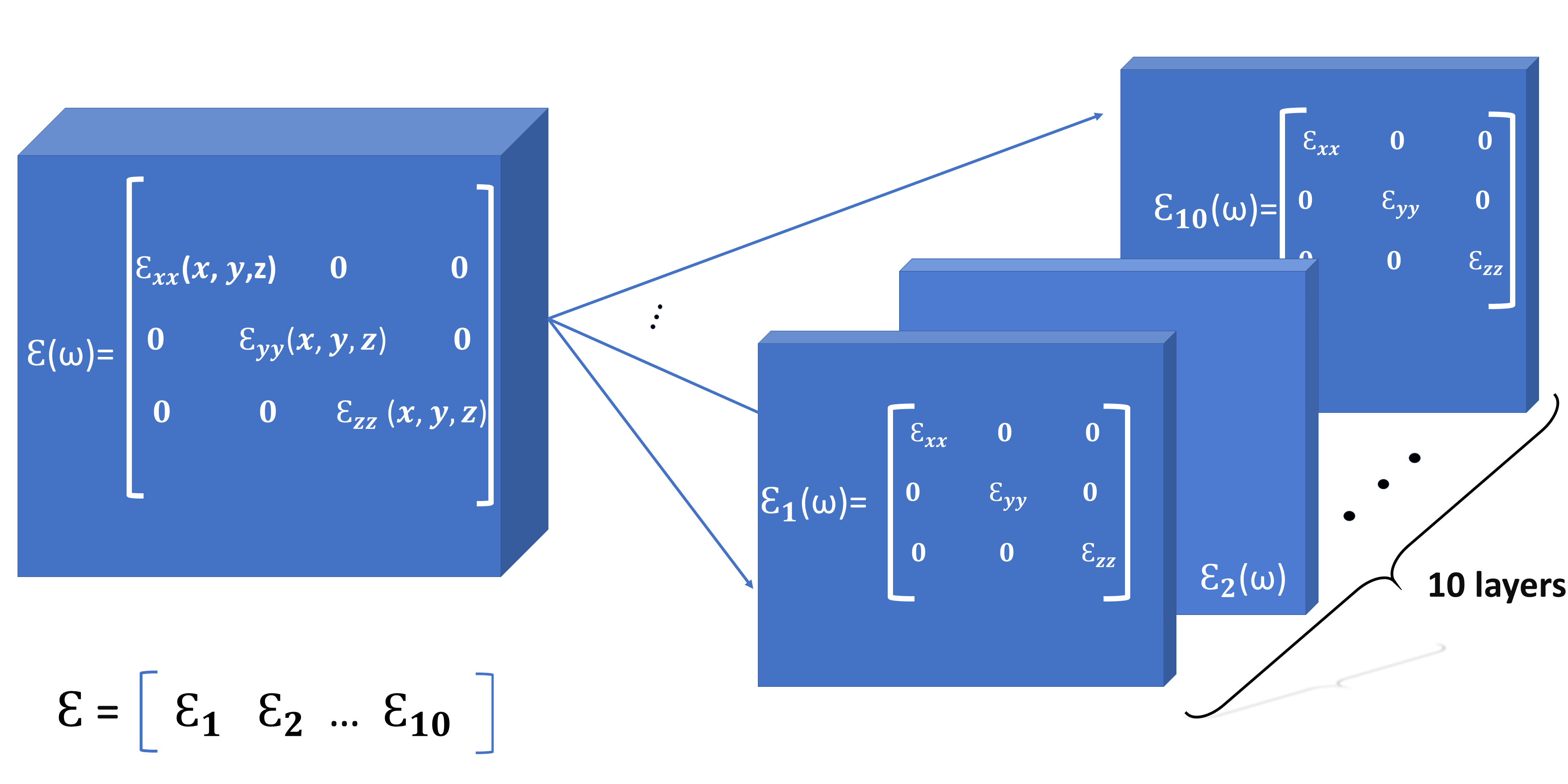}
	\caption{Overview of  inhomogeneous-anisotropic wall } 
\end{figure}

We modeled the wall with airgap using three layers, placing two layers of homogeneous wall with $\varepsilon_{r}= 6$ on either side and in the center of an air layer, which is a practical example of an electromagnetic complex wall.

\subsection{Deep Learning}

Artificial neural networks, which are inspired by brain neural cells, emerged in the last two decades and have a wide range of applications, including optimization, artificial intelligence, and many others. By modeling the structure of a cell and the neural network of the brain, an artificial cell and neural network can be constructed.
An overview of an artificial neuron and neural cell are depicted in Figure 4. The inputs (input neurons) are as follows: $X_{1}$, $X_{2}$, ... . In a neural network, each $X_{i}$ has a weight, denoted by $W_{i}$. Indeed, each input is multiplied by the weight associated with it. The sum function (sigma) in the neural network then adds the product of the X's and W's, and an activation function  then calculates the output value based on this computation. If $f(u)$ represents the activation function and $b$ represents a bias value, the output of neurons can be written as follows:

\begin{figure}[h]
	\centering
	\includegraphics[scale=0.08]{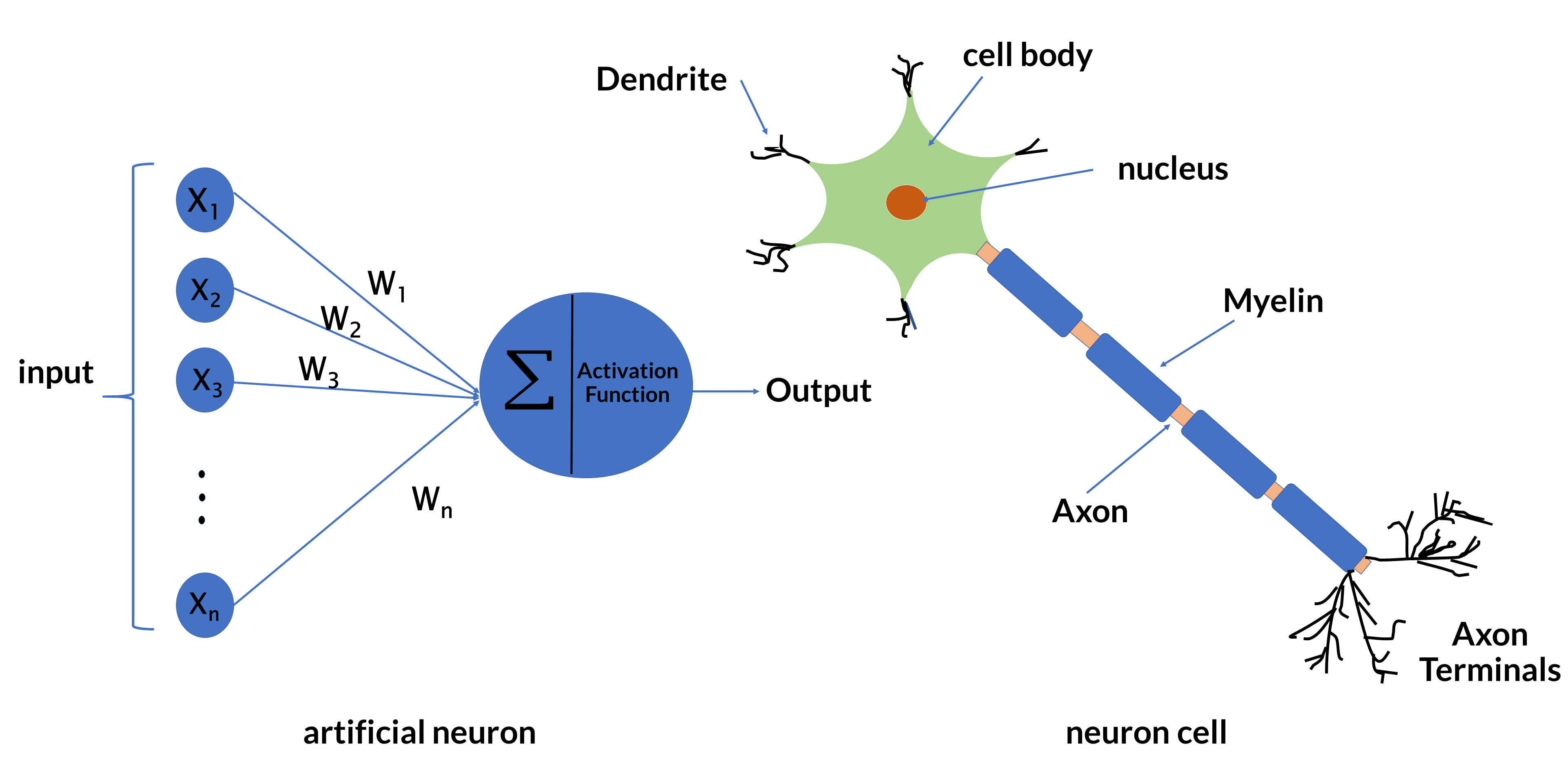}
	\caption{An overview of an artificial neuron and neural cell} 
\end{figure}

\begin{equation}
Y =f(\sum\limits_{i=1}^n W_{i}X_{i}+b_{i})
\end{equation}

A neural network is composed of multiple layers of neurons. In general, a neural network consists of three layers: input, hidden, and output. As the number of layers and neurons in each hidden layer increases, the model becomes increasingly complex. As the number of hidden layers and neurons in our network increases, it becomes a deep neural network, and learning is referred to as deep learning. The activation of Relu and Linear functions was used to design the deep neural network, as shown in Equations 8 and 9.

\begin{equation}
f(x) = \begin{cases}
0 & x \leq 0 \\
x & x > 0
\end{cases} 
\end{equation}

\begin{equation}
f(x) = x
\end{equation}

\subsection{FDTD}
The electromagnetic solver in the finite difference time domain (FDTD) is one technique for solving Maxwell equations. Ampere's and Faraday's Laws can be expressed as the following equations:
\begin{equation}
\begin{aligned}
\quad\nabla\times{H} = \frac{\partial{D}}{\partial t} = \varepsilon \frac{\partial{E}}{\partial t} \quad & ,   & \quad \text{(Faraday's Law)} \\[5pt]
\quad\nabla\times{E} = -\frac{\partial{B}}{\partial t} = \mu \frac{\partial{H}}{\partial t} \quad & ,  & \quad \text{(Ampere's Law)}   \\[5pt]
\end{aligned}
\end{equation}
$\varepsilon$ and $\mu$ denote the relative permittivity and permeability, respectively. They can be scalar, tensor, or location dependent, as explained previously. We employ  $TM_{z} $ polarization in this work in order to rewrite Ampere's and Faraday's Laws as follows\cite{schneider2010understanding}:

\begin{equation}
\epsilon\frac{\partial{E}}{\partial t} =\quad\nabla\times{H} = \begin{vmatrix}
\hat{a_{x}} & \hat{a_{y}} & \hat{a_{z}} \\ 
\frac{\partial}{\partial x} & \frac{\partial}{\partial y} & 0\\ 
H_{x} & H_{y} & 0\\ 
\end{vmatrix} = \hat{a_{z}}(\frac{\partial H_{y}}{\partial x}-\frac{\partial H_{x}}{\partial y})
\end{equation}

\begin{equation}
-\mu\frac{\partial{H}}{\partial t} =\quad\nabla\times{E} = \begin{vmatrix}
	\hat{a_{x}} & \hat{a_{y}} & \hat{a_{z}} \\ 
	\frac{\partial}{\partial x} & \frac{\partial}{\partial y} & 0\\ 
	0 & 0 & E_{z}\\ 
\end{vmatrix} = \hat{a_{x}}\frac{\partial E_{z}}{\partial y}-\hat{a_{y}}\frac{\partial E_{z}}{\partial x}
\end{equation}

The scalar equations for $TM_{z} $ are obtained from (11) and (12):

\begin{equation}
-\mu\frac{\partial{H_{x}}}{\partial t} = \frac{\partial E_{z}}{\partial y}
\end{equation}

\begin{equation}
\mu\frac{\partial{H_{y}}}{\partial t} = \frac{\partial E_{z}}{\partial x}
\end{equation}

\begin{equation}
\epsilon\frac{\partial{E_{z}}}{\partial t} = \frac{\partial H_{y}}{\partial x}-\frac{\partial H_{x}}{\partial y}
\end{equation}

Equations (13)–(15) can be expressed in finite-differences form due to the discretization in space-time; future fields can be written in terms of past fields. Assume that the indexes $m$ and $n$ denote the spatial step in the $x$ and $y$ direction, and $q$ denotes the temporal step. Additionally, the spatial step sizes are $\Delta x$ and $\Delta y$ in the $x$ and $y$ directions, respectively.
The finite difference approximation of (13) expanded about
the space-time point (m$\Delta x$, (n + 1/2)$\Delta y$, q$\Delta t$).The resulting equation is:

\begin{align}
\begin{split}
-\mu\frac{H_x^{q+\frac{1}{2}}[m,n+\frac{1}{2}] - H_x^{q-\frac{1}{2}}[m,n+\frac{1}{2}]}{\Delta t}=\\
\frac{E_z^{q}[m,n+1] - E_z^{q}[m,n]}{\Delta y}
\end{split}
\end{align}

For future value in terms of past value, the equation can be rewritten as follows:
\begin{align}
\begin{split}
H_x^{q+\frac{1}{2}}[m,n+\frac{1}{2}] = H_x^{q-\frac{1}{2}}[m,n+\frac{1}{2}]-\\ \frac{\Delta t}{\mu \Delta y} (E_z^{q}[m,n+1] - E_z^{q}[m,n])
\end{split}
\end{align}

We can also write for Equation (14) and (15) expanded about
the space-time point ((m + 1/2)$\Delta x$, n$\Delta y$, q$\Delta t$) and (m$\Delta x$, n$\Delta y$, (q + 1/2)$\Delta t$), respectively:

\begin{align}
\begin{split}
H_y^{q+\frac{1}{2}}[m+\frac{1}{2},n] = H_y^{q-\frac{1}{2}}[m+\frac{1}{2},n]+\\ \frac{\Delta t}{\mu \Delta x} (E_z^{q}[m+1,n] - E_z^{q}[m,n])
\end{split}
\end{align}

\begin{align}
\begin{split}
E_z^{q+1}[m,n] = E_z^{q}[m,n] + \frac{\Delta t}{\epsilon \Delta x} \lbrace H_y^{q+\frac{1}{2}}[m+\frac{1}{2},n] - \\
H_y^{q-\frac{1}{2}}[m-\frac{1}{2},n] \rbrace - \frac{\Delta t}{\epsilon \Delta y}v\lbrace H_y^{q+\frac{1}{2}}[m,n+\frac{1}{2}]\\ -H_y^{q+\frac{1}{2}}[m,n-\frac{1}{2}]  \rbrace
\end{split}
\end{align}

In the simulation, we consider a two-dimensional environment on the X-Y plane with dimensions of 2 x 2 meters, surrounded by perfectly matched layers. The spatial step sizes are set to be $0.1 \lambda$ in the X and Y directions .
Given that we have assumed a frequency of 3 GHz, $\lambda = c/f= 0.1$m. This means that the size of the meshes is 0.01m. Given a space of 2 x 2 meters, the number of meshes in the X and Y directions is 200.

\section{Data Gathering}
In a two-dimensional through-the-wall radar problem, all materials are invariant within the z-direction. We performed the simulations using the FDTD technique. The simulation is carried out using the FDTD library in Python \cite{fdtdpythonlib}. We consider square targets with a permittivity of $\varepsilon_{r} = 80$ and dimensions of 10, 20, and 30 cm; this permittivity value is comparable to that of water, allowing us to estimate the target in a manner analogous to that of the human body. Additionally, we considered a wall thickness of 10 cm. However, in the case of a wall with an air gap, we used a thickness of 15 cm to account for the presence of a 5 cm layer of air within. We generated a plane wave with a frequency of 3 GHz using the FDTD library's line source. $E_{x} = E_{y} = H_{z} = 0$ and $E_{z} , H_{x} , H_{y} $  are non-zero for $TM_{z} $ polarization. When the source's wave reaches the wall, some of it returns, while the remainder passes through, reaches the target, and scatters. The detector collects the scattered wave. The $E_{z}, H_{x}, and H_{y} $ fields are obtained and used to construct the required dataset. We generate a dataset by moving the target or targets behind the wall in the coordinates X=[5,85] and Y=[40,100]. When there is only one target, we move it in the specified coordinates with a step of 10 cm in both the x and y directions, and we record the coordinates of the target center in the scattered fields in the database for each shift. When two targets exist, we proceed in the same manner as before, except that duplicate states are eliminated. The target location is the two-dimensional coordinates of the target's center. The built-in database stores the two-dimensional coordinates of targets with contiguous squared fields. Figure 5 shows an overview of the problem, as well as the electric and magnetic fields associated with the homogeneous wall with two targets.

\begin{figure*}[h]
	\centering
	\subfloat[]{\includegraphics[width=6cm]{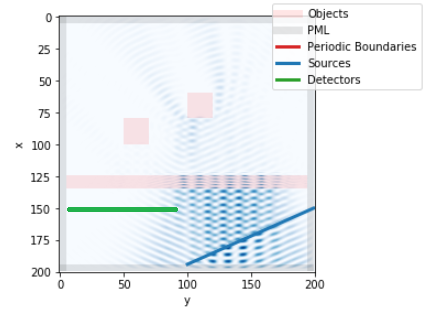}}
	\qquad
	\subfloat[]{\includegraphics[width=6cm]{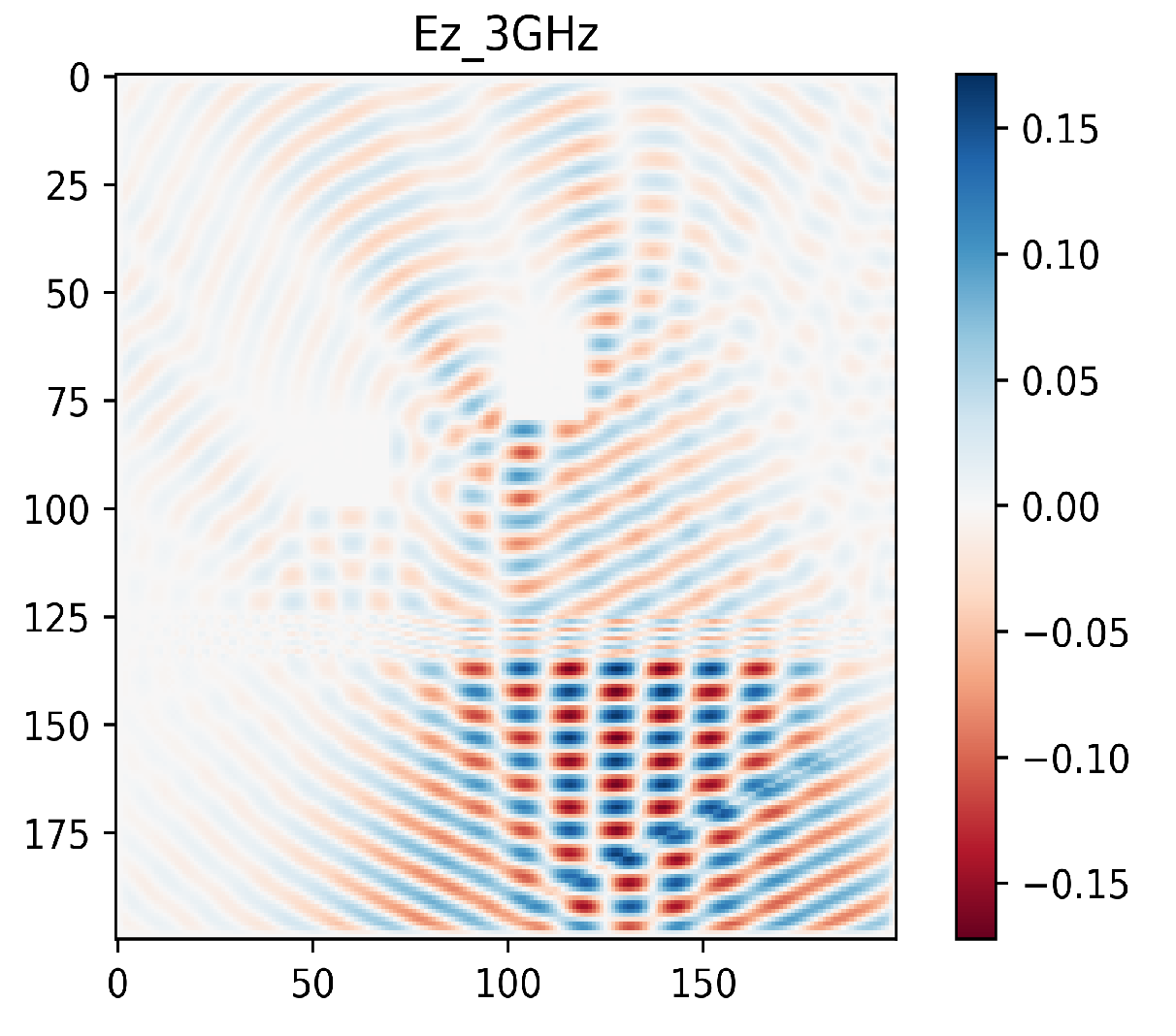}}
	\qquad
	\subfloat[]{\includegraphics[width=6cm]{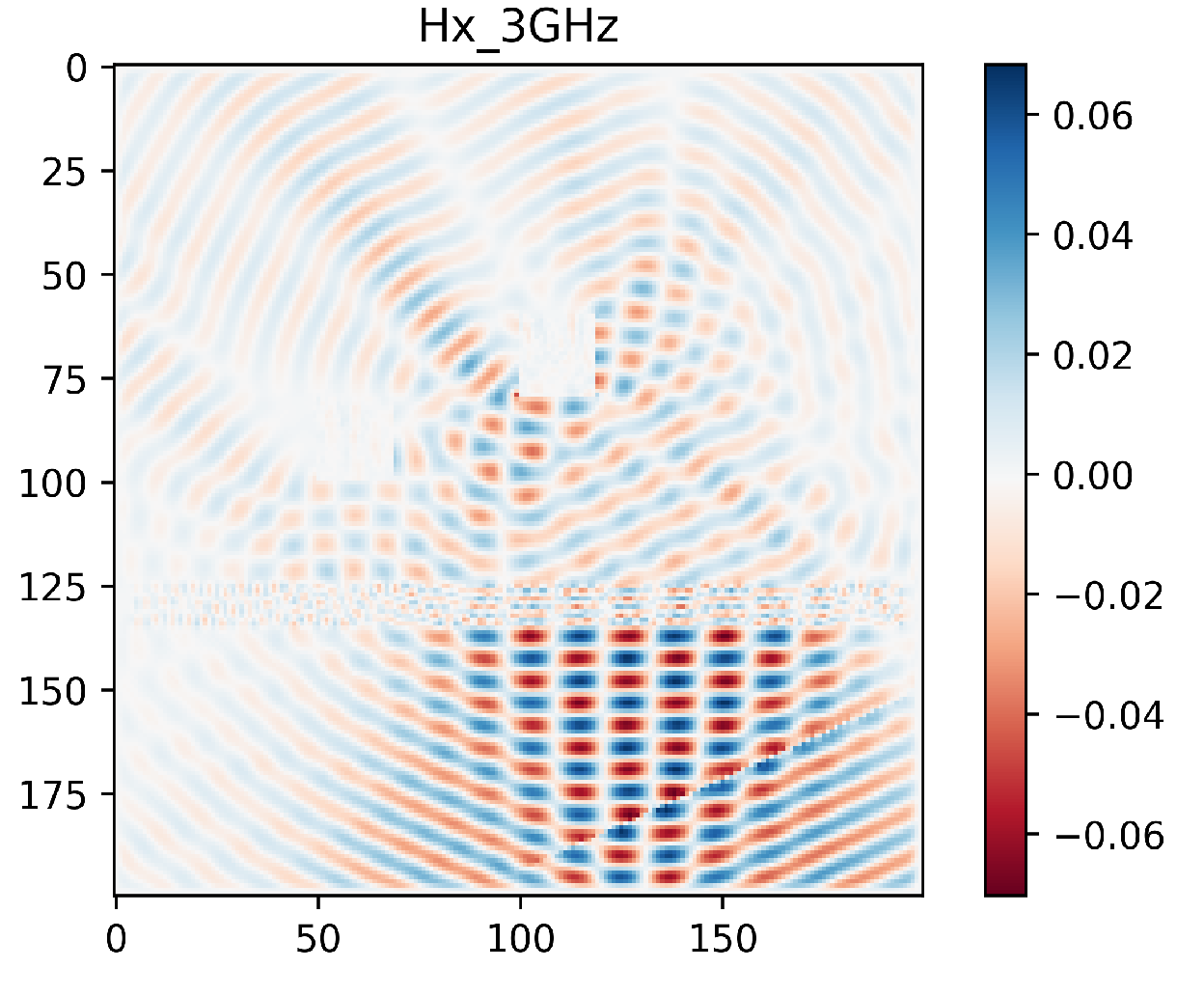}}
	\qquad
	\subfloat[]{\includegraphics[width=6cm]{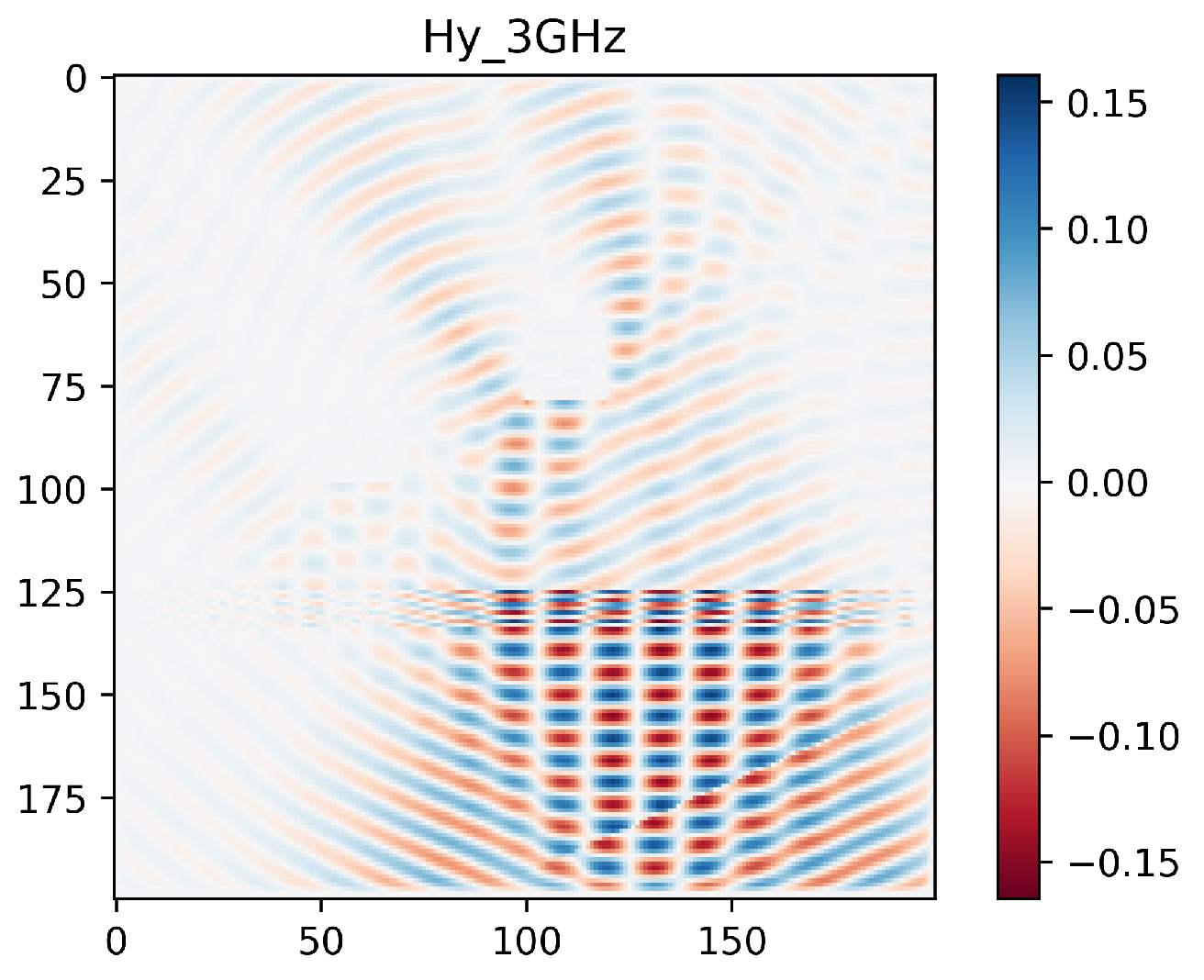}}
	\qquad
	\caption{fig (a) an overview of the problem. Figs (b), (c), (d) $E_{z}$, $H_{x}$, $H_{y}$ in plane z = 0}
	\label{4fig}
\end{figure*}

We generate 63 data points for each wall scenario in the single-target mode and 756 data points for the two-target mode (this is not the number of all possible data). As we have five different wall scenarios, the number of datasets is multiplied by five. Thus, if all data is provided to the model, we have 315 data points for the single-target mode and 3780 data points for the two-target mode. In each of these cases, we allocate $70\%$ of the data to the train, $20\%$ to the test, and $10\%$ to validation.

\section{NUMERICAL AND EXPERIMENTAL RESULTS}

\subsection{without noise}

We used a Deep Neural Network (DNN) in this work to estimate the two-dimensional position of targets hidden behind a wall. The DNN algorithm is implemented in Python, and the model is constructed using the Tensorflow and Keras \cite{chollet2015keras} frameworks.We used dense and dropout layers in the proposed deep neural network model. The fully connected (dense) layer contains all of the neurons from the previous layer. The fully connected layer's primary function is to connect the lower layer's local features to the upper layer's local features. Several neurons in the dropout layer are inadvertently ignored during the learning process. Dropout is a regularization technique for neural networks that assists in avoiding the learning process from being manipulated, as well as increasing learning speed and reducing the risk of overfitting. During the training process, dropout should be applied to the connections between layers, but not during network evaluation and testing. We propose a distinct deep learning model for each of the single-target and two-target locating modes that performs well regardless of the assumed wall models. Additionally, we investigated two distinct modes: one without noise, which corresponds to an infinite SNR, and one with noise at various SNRs.

We employed a 9-layer deep neural network to locate a sngle target. In this case, the DNN input's size is 285, which is equal to the length of the $E_{z}$, $H_{x}$, and $H_{y}$ field vectors, which are composed of three vectors with a length of 95. The DNN's output has a size of 2, estimating the two numbers that represent the target's x and y coordinates. We chose 0.0001 as the learning rate and 30 as the batch size. Additionally, we used the Adam optimizer and the Mean Squared Logarithmic Error (MSLE) loss function, which are defined below:
\begin{align}
L(y,\hat{y}) = \frac{1}{N} \displaystyle\sum\limits_{i=0}^N (\log(y_{i}+1)-\log(\hat{y_{i}}+1))^2
\end{align}
y is the actual value, $\hat{y}$ is estimated value, and $N$ is the total number of data points. In fact, MSLE is the mean of the squared differences between the actual and estimated values following log transformation. The details of the designed DNN for this mode are summarized in Table \Romannum{2}.

\renewcommand{\arraystretch}{1.5}
\begin{table}[h]
	\centering
	\caption{DNN architecture used for location for single-target mode}
			\begin{tabular}{|l|l|l|l|l|}
				\hline
				\textbf{layer number} & \textbf{layer}                & \textbf{output shape} & \textbf{number of parameters} & \textbf{activation function} \\ \hline
				1            & dense\_1 (Dense)     & (,284)  & 80940                & relu                \\ \hline
				2            & dropout\_1 (Dropout) & (,284)  & 0                    & -                   \\ \hline
				3            & dense\_2 (Dense)     & (,300)  & 85500                & relu                \\ \hline
				4            & dropout\_2 (Dropout) & (,300)  & 0                    & -                   \\ \hline
				5            & dense\_3 (Dense)     & (,300)  & 90300                & relu                \\ \hline
				6            & dropout\_3 (Dropout) & (,300)  & 0                    & -                   \\ \hline
				7            & dense\_4 (Dense)     & (,300)  & 90300                & relu                \\ \hline
				8            & dropout\_4 (Dropout) & (,300)  & 0                    & -                   \\ \hline
				9            & dense\_5 (Dense)     & (,2)    & 602                  & linear              \\ \hline
			\end{tabular}
\end{table}

We combined the Dense and Dropout layers in the presented model through trial and error and experience to achieve high accuracy and low error. We attempted to achieve the highest accuracy possible while using the smallest model possible. The accuracy and error graphs for single-target locating over 5000 epochs are shown in Figures 6.(a) and 6.(b).
Additionally, we used a 12-layer deep neural network to locate two targets. The accuracy and error graphs for two targets over 1000 epochs are shown in Figure 6.(c), 6.(d), 6.(e), and 6.(f). In these two modes, we used the Mean Squared Logarithmic Error (MSLE) loss function and the Adam optimizer, with the learning rate set to 0.001 and the batch size set to 30. The details of the designed networks are summarized in Table \Romannum{3}, \Romannum{4}. 

\renewcommand{\arraystretch}{1.5}
\begin{table}[h]
	\centering
	\caption{DNN architecture used for location for two-targets mode}
			\begin{tabular}{|l|l|l|l|l|}
				\hline
				\textbf{layer number} &\textbf{ layer}                & \textbf{output shape} & \textbf{number of parameters}& \textbf{activation function} \\ \hline
				1            & dense\_1 (Dense)     & (,285)  & 81510                & relu                \\ \hline
				2            & dropout\_1 (Dropout) & (,285)  & 0                    & -                   \\ \hline
				3            & dense\_2 (Dense)     & (,300)  & 85800                & relu                \\ \hline
				4            & dropout\_3 (Dropout) & (,300)  & 0                    & -                   \\ \hline
				5            & dense\_3 (Dense)     & (,300)  & 90300                & relu                \\ \hline
				6            & dropout\_4 (Dropout) & (,300)  & 0                    & -                   \\ \hline
				7            & dense\_4 (Dense)     & (,300)  & 90300                & relu                \\ \hline
				8            & dropout\_5 (Dropout) & (,300)  & 0                    & -                   \\ \hline
				9            & dense\_5 (Dense)     & (,300)  & 90300                & relu                \\ \hline
				10           & dropout\_6 (Dropout) & (,300)  & 0                    & -                   \\ \hline
				11           & dense\_6 (Dense)     & (,300)  & 90300                & relu                \\ \hline
				12           & dense\_7 (Dense)     & (,4)    & 1204                 & linear 					\\ \hline
			\end{tabular}
\end{table}

Because the received signal has not changed, the DNN's input in the two-targets mode is identical to the single-target mode. However, in two-target mode, we must estimate the two targets' two-dimensional coordinates, resulting in a network output of four neurons.
When we compare the network parameters, also known as trainable parameters, for single- and two-target modes, as shown in Tables \Romannum{2}, \Romannum{3}, and \Romannum{4}, we find that single-target mode has 347,642 parameters and two-target mode has 529,714 parameters, indicating that the more targets we have, the more parameters we have. This results in a more dense network, which can be achieved by increasing the number of neural network layers or by increasing the number of neurons contained within the layers.

The accuracy, validation-accuracy, loss, and validation-loss values for the five models considered for the wall, as well as a model incorporating all data in single and double locating, are shown in Table \Romannum{5}. The proposed deep learning model for location finding has been fine-tuned in such a way that it is insensitive to changing the wall model and does not change the amount of accuracy and error associated with changing the wall model. As can be seen, the problem becomes more complicated as the number of targets increases, and accuracy decreases as the network becomes more dense.

\begin{figure*}[th]
	\centering
	\subfloat[single target accuracy]{\includegraphics[width=6cm]{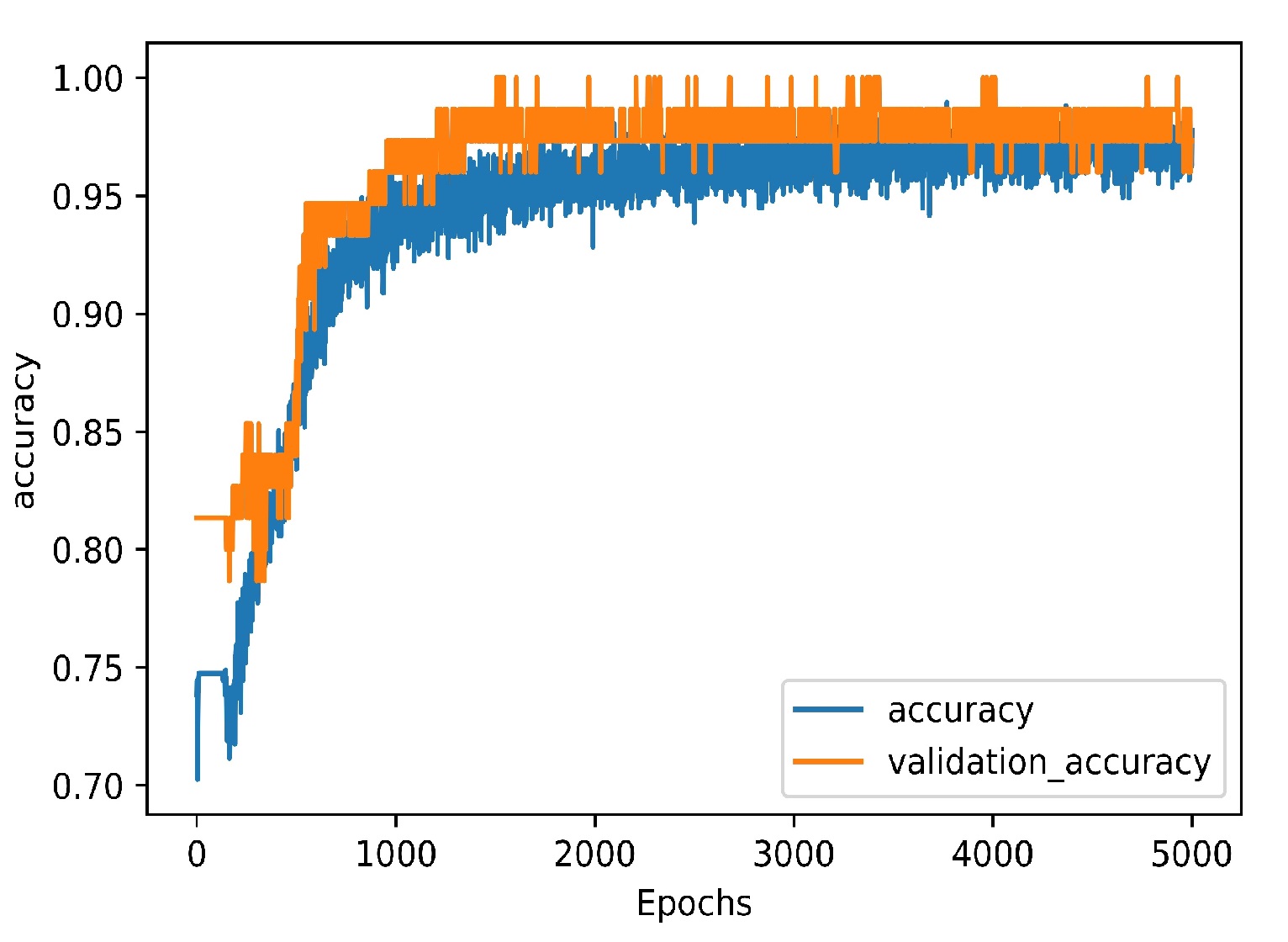}}
	\qquad
	\subfloat[single target loss]{\includegraphics[width=6.3cm]{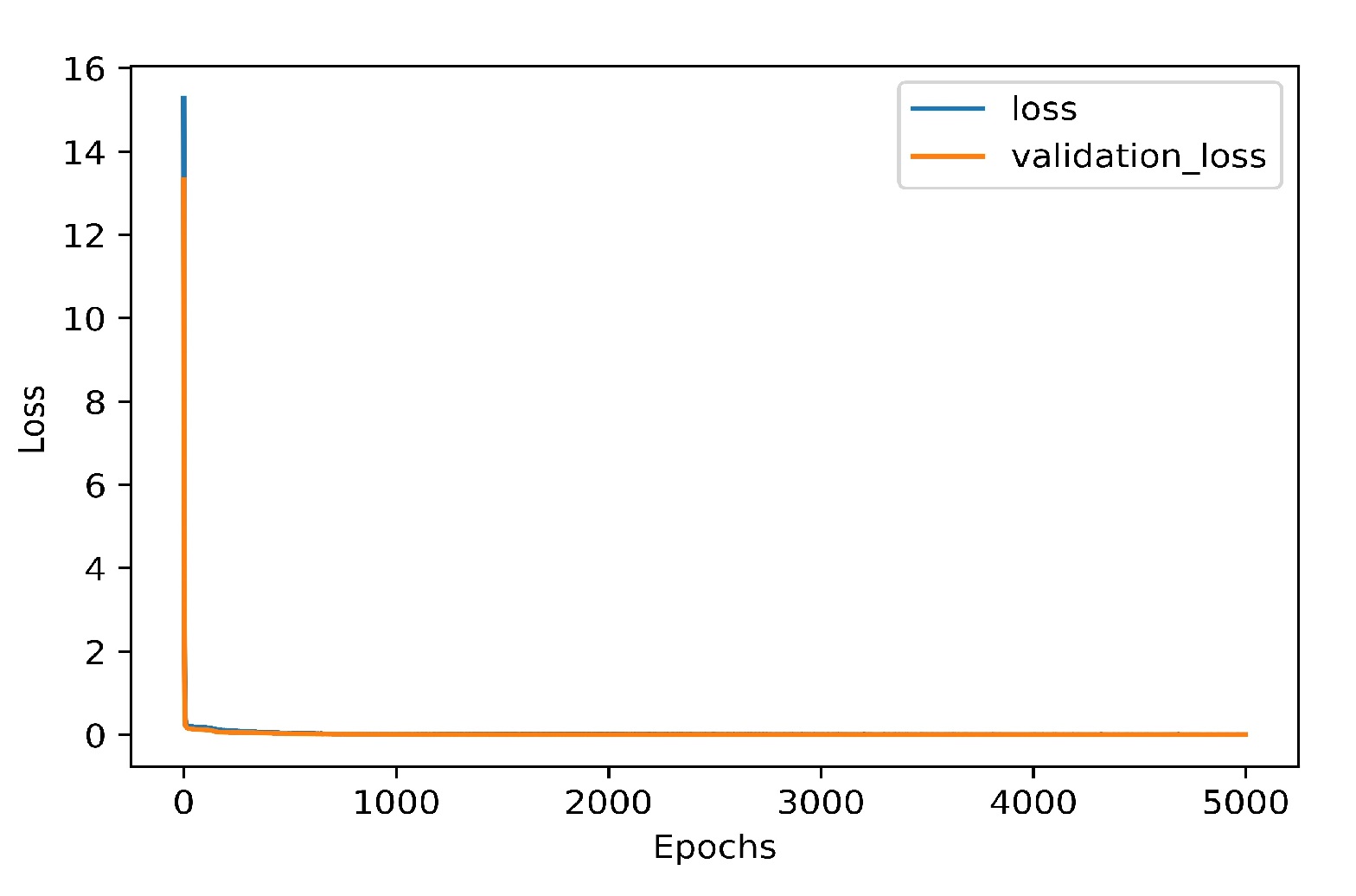}}
	\qquad
	\subfloat[two targets accuracy]{\includegraphics[width=6cm]{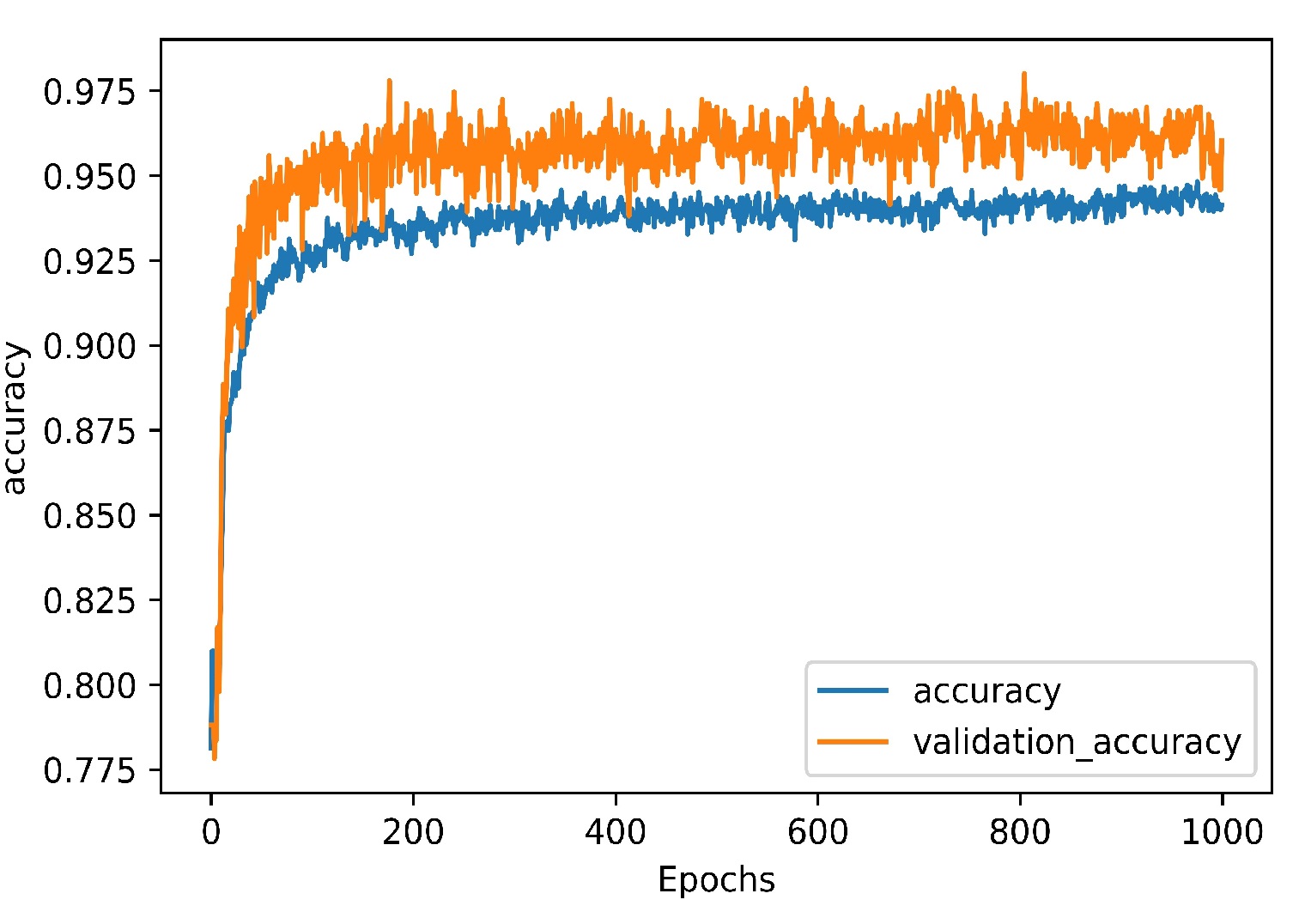}}
	\qquad
	\subfloat[two targets loss]{\includegraphics[width=6cm]{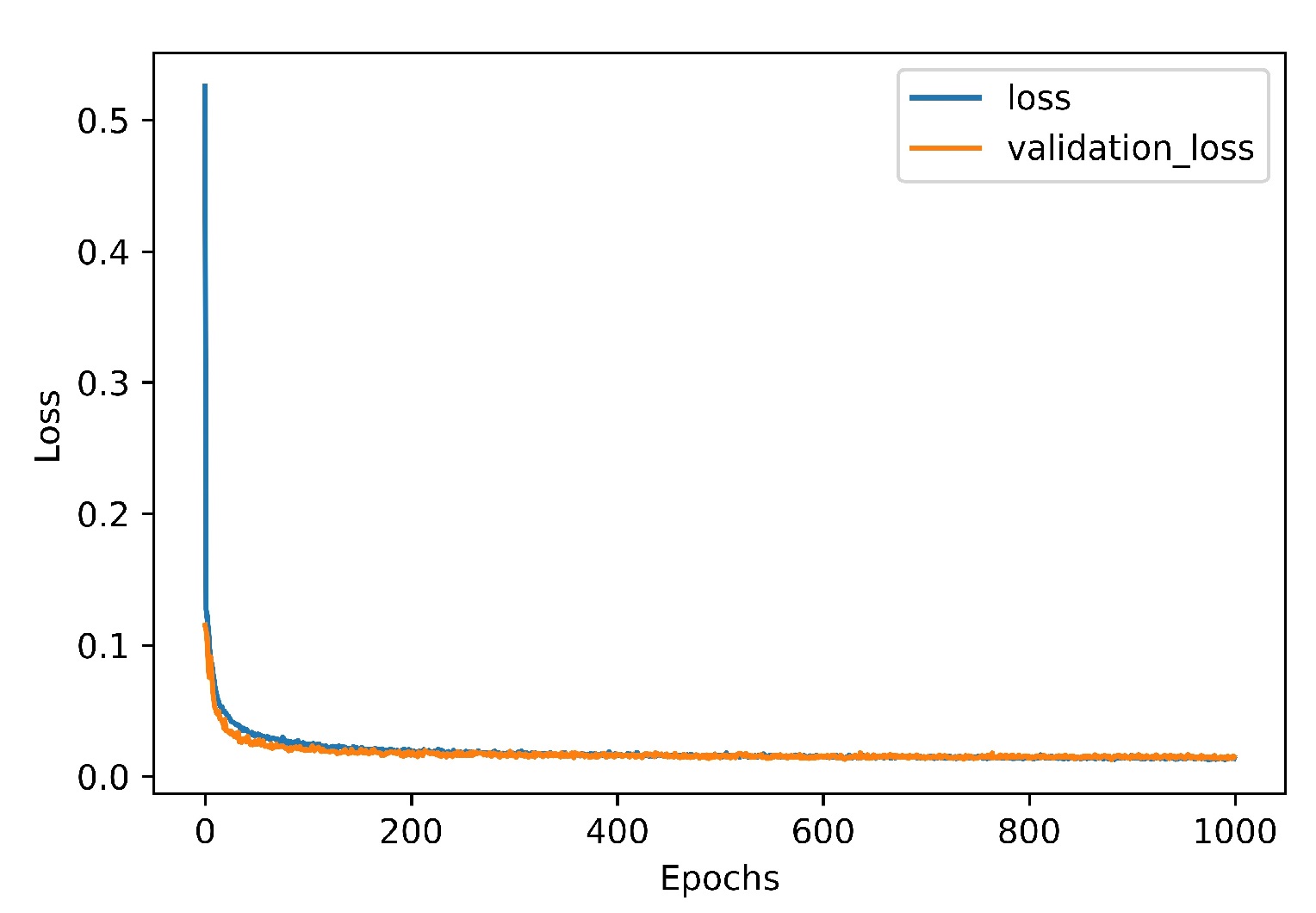}}
	\qquad
	\caption{Accuracy and loss diagrams in single and two targets locating for all data.}
	\label{6fig}
\end{figure*}

\renewcommand{\arraystretch}{1.4}
\begin{table*}[h]
	\centering
	\caption{The value of accuracy, validation-Accuracy, loss and validation-loss for the 5 wall models and a model containing all data assumed in single, double and triple locating for without noise mode}
	\begin{tabular}{ll|l|l|l|l|l|l|}
		\cline{3-8}
		&                      &\textbf{ homogeneous} & \textbf{airgap} &\textbf{ inhomogeneous}& \textbf{anisotropic} & \textbf{inhomogeneous-anisotropic} & \textbf{all data} \\ \hline
		\multicolumn{1}{|c|}{\multirow{4}{*}{\begin{tabular}[c]{@{}c@{}}\textbf{Single}\\ \textbf{Target}\end{tabular}}} & Accuracy             & \%97        & \%96.9 & \%96.9        & \%94        & \%96                      & \%97.7   \\ \cline{2-8} 
		\multicolumn{1}{|c|}{}                                                                            & Loss                 & 0.001       & 0.009  & 0.007         & 0.007       & 0.008                     & 0.005    \\ \cline{2-8} 
		\multicolumn{1}{|c|}{}                                                                            & validation\_Accuracy & \%93        & \%100  & \%100         & \%93.3      & \%100                     & \%97.3   \\ \cline{2-8} 
		\multicolumn{1}{|c|}{}                                                                            & validation-loss      & 0.09        & 0.07   & 0.01          & 0.01        & 0.008                     & 0.002    \\ \hline
		\multicolumn{1}{|l|}{\multirow{4}{*}{\begin{tabular}[c]{@{}l@{}}\textbf{Two} \\ \textbf{Targets}\end{tabular}}}     & Accuracy             & \%95.6      & \%94.7 & \%96          & \%95        & \%96.2                    & \%94.1   \\ \cline{2-8} 
		\multicolumn{1}{|l|}{}                                                                            & Loss                 & 0.008       & 0.009  & 0.01          & 0.007       & 0.006                     & 0.01     \\ \cline{2-8} 
		\multicolumn{1}{|l|}{}                                                                            & validation-Accuracy  & \%96.5      & \%95   & \%95.1        & \%94.1      & \%96.5                    & \%94.5   \\ \cline{2-8} 
		\multicolumn{1}{|l|}{}                                                                            & validation-loss      & 0.01        & 0.01   & 0.02          & 0.01        & 0.01                      & 0.01     \\ \hline
		
	\end{tabular}
\end{table*}

\subsection{With noise}

Additionally, adding noise to the signal can help bring the situation closer to reality. In this section, we added noise to the received signal and assumed the wall is a complex electromagnetic wall, increasing the difficulty of the task. To evaluate the model's performance in the presence of noise, we introduced an Additive White Gaussian Noise (AWGN) with varying SNR values. The accuracy value for the validation dataset in all data mode is shown in Figure 7.
\begin{figure}[h]
	\centering
	\includegraphics[scale=0.47]{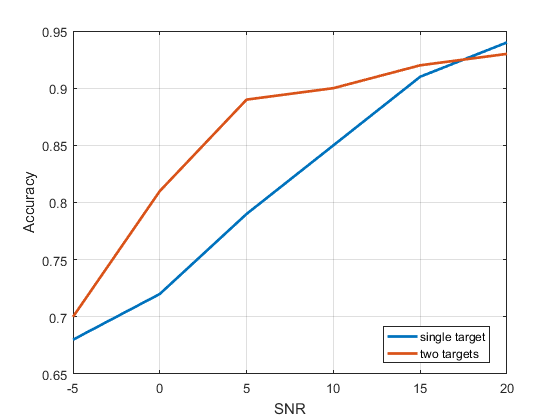}
	\caption{Comparison of accuracy on validation datasets for all two single-target and two-targets modes for different SNRs } 
\end{figure}
We made no changes to the structure of the neural network models presented in this scenario for single-target and two-target modes.
As illustrated in Figure 8, accuracy increases as SNR increases, but the point is that when single-target and two-target diagrams are compared. Single-target mode is less accurate at low SNRs, whereas it is more accurate at high SNRs. To clarify this point, one could argue that because single-target mode has fewer datasets than two-target mode, the neural network would overfit after several epochs, which is why we use early stopping to avoid overfitting. The training procedure is terminated before the neural network becomes overfit. Due to a lack of data in the single-target mode, we have lower accuracy in low SNRs compared to the two-target mode. However, as the SNR increases, the neural network overcomes this situation and achieves greater precision than the two-target mode.

\section{CONCLUSIONS}

We performed TWR multi-targets two-dimensional locating in this paper using several models that are close to a realistic wall model. We considered five distinct wall models: homogeneous, with an air gap, inhomogeneous, anisotropic, and inhomogeneous-anisotropic. Additionally, we evaluated three distinct target sizes (10, 20, and 30 cm) and two distinct target numbers (single and double). We used two deep learning models for the two modes (single and double targets). Each model can locate targets in the three sizes mentioned above under the various wall conditions. Assuming infinite SNR (i.e., no noise), we achieved $97.7\%$ accuracy on all data for single-target two-dimensional locating, which includes all assumed wall models and all targets in three different sizes. We were able to evaluate the models' performance at various SNR levels by introducing noise into them. Low SNRs resulted in a $10\%–20\%$ reduction in accuracy, while high SNRs resulted in a less than $10\%$ reduction in accuracy.
\\

\appendices

\ifCLASSOPTIONcaptionsoff
  \newpage
\fi

\bibliographystyle{IEEEtran}
\bibliography{myref}

\end{document}